\documentclass[runningheads]{svmult}

\usepackage{makeidx}   
\usepackage{graphicx}  
\usepackage{subeqnar}  
\usepackage{multicol}  
\usepackage{physprbb}  
\makeindex             

\newcommand{\Lya}{Ly$\alpha\ $}

\def\kmsmpc{\,{\rm km\,s^{-1}\,Mpc^{-1}}}
\def\kms{\,{\rm km\,s^{-1}}}

\newcommand{\etal}{et~al.\ }

\def\msun{\,{\rm M_\odot}}

\def\HI{\hbox{H~$\scriptstyle\rm I\ $}}
\def\HII{\hbox{H~$\scriptstyle\rm II\ $}}

\def\spose#1{\hbox to 0pt{#1\hss}}
\def\lta{\mathrel{\spose{\lower 3pt\hbox{$\mathchar"218$}} \raise 2.0pt\hbox{$\mathchar"13C$}}}
\def\gta{\mathrel{\spose{\lower 3pt\hbox{$\mathchar"218$}} \raise 2.0pt\hbox{$\mathchar"13E$}}}
\def\lya{Ly$\alpha\ $}

\def\sfr{\,{\rm M_\odot\,yr^{-1}}}

\begin{document}
\title*{First Light and the Reionization of the Universe\footnote{Invited
talk at the ESO-CERN-ESA Symposium on Astronomy, Cosmology, and Fundamental 
Physics, March 4-7 2002, Garching, Germany.}}
\toctitle{First Light and the Reionization of the Universe}
\titlerunning{First Light}
%
\author{Piero Madau}
\authorrunning{Piero Madau}
%
%
\institute{University of California Santa Cruz, CA 95064, USA}

\maketitle              

\begin{abstract}
The development of primordial inhomogeneities into the non-linear regime
and the formation of the first bound objects mark the transition from 
a simple cooling universe -- described by just a few parameters -- to a 
very messy hot one -- the realm of radiative, hydrodynamic and star 
formation processes.
In popular cold dark matter cosmological scenarios, stars and quasars 
may have first appeared in significant numbers around a redshift of 10 
or so, as the gas within 
subgalactic halos cooled rapidly due to atomic processes and fragmented.
This early generation of sources generated the ultraviolet radiation and 
kinetic energy that ended 
the cosmic ``dark ages'' and reheated and reionized most of the hydrogen in 
the cosmos by a redshift of 6. The detailed thermal history of the 
intergalactic medium -- the main repository of baryons at high redshift --
during 
and soon after these crucial formative stages depends on the power-spectrum 
of density fluctuations on small scales and on a complex network of poorly 
understood `feedback' mechanisms, and is one of the missing link in galaxy 
formation and evolution studies. 

\end{abstract}

\section{Introduction}

The past few years have witnessed great progress in our understanding of the
high-redshift universe. The pace of observational cosmology and extragalactic
astronomy has never been faster. {\it Hubble Space Telescope} deep 
imaging surveys and 
spectroscopic follow-up from 8--10 m telescopes are together elucidating
the history of star formation and galaxy clustering back to redshift $z=4$.
The measurement of the far-IR/sub-mm background by DIRBE and FIRAS onboard
the {\it COBE} satellite and the detection of distant ultraluminous 
sub-mm sources by the SCUBA camera have shed new light on the 
`optically hidden' side of galaxy formation, and shown that a significant 
fraction of the energy released by stellar nucleosynthesis is re-emitted
as thermal radiation by dust. Ongoing X-ray studies with the {\it Chandra} 
and {\it XMM Newton} satellites may be discovering 
the population of highly absorbed quasistellar objects predicted to 
be responsible for the hard X-ray background. Big surveys currently in 
progress such as the Sloan Digital Sky Survey, together with the use 
of novel instruments have led to the discovery of 
galaxies and quasars at redshifts $z\gta 6$ \cite{fan} (the current record 
holder appears to be a 
$z=6.56$ galaxy gravitationally lensed by a foreground cluster \cite{hu02}),
when the universe was about 6\% of its current age. These observations 
may be at the threshold of probing the epoch of first light and 
cosmological reionization.
{\it Keck} and {\it VLT} observations of redshifted \HI \Lya (`forest') 
absorption in the spectra of distant quasars are becoming an increasingly
sensitive probe of the distribution of gaseous matter in the universe, and
are revealing the topology, thermal and ionization state, and chemical
composition of the intergalactic medium (IGM) -- the main repository of
baryons at high redshift -- in unprecedented detail.
From all these data, an empirical picture is beginning to emerge
of the processes and timescales whereby ordinary matter in the
universe has been transformed as it is subjected to radiative and chemical
interactions and processed through stars.

On the theoretical side progress has been equally significant. The key 
idea, that primordial density fluctuations grow by gravitational instability 
driven by cold dark matter (CDM), 
has been elaborated upon and explored in detail through large-scale 
numerical simulations on supercomputers, leading to a hierarchical 
(`bottom-up') scenario of structure formation. In this model, the first 
objects to form are on subgalactic scales, and merge to make progressively 
bigger structures (`hierarchical clustering'). Ordinary matter in the 
universe follows the dynamics 
dictated by the dark matter until radiative, hydrodynamic, and star formation 
processes take over. Perhaps the most remarkable success of this theory has 
been the prediction of anisotropies in the temperature of the cosmic microwave
background (CMB) radiation at about the level subsequently measured by the
{\it COBE} satellite and most recently by the {\it BOOMERANG}, {\it MAXIMA},
{\it DASI}, and {\it CBI} experiment.
 
In spite of the remarkable progress achieved in recent years, many
fundamental questions (beside, of course, the nature of the dark matter
and dark energy components) remain only partially answered. How
and when was the universe reheated? We know that at least some galaxies
and quasars were already shining when the universe was less than $10^9\,$ yr
old. But when exactly did the first luminous structures form and how bright
were they? We believe 
there is a strong coupling between the thermodynamic state of the IGM
and the process of galaxy formation, and we suspect complex feedback
mechanisms are at work in this interaction. The detailed astrophysics of
these processes is, however, unclear. Finally, the precise location 
and degree of metal enrichment of most the baryons in the universe 
remains an open question.

\section{The dark ages}

At epochs corresponding to $z\sim 1200$, about half a million years after 
the big bang, the universe recombined, became optically thin to Thomson 
scattering, and entered, in the words of Sir Martin Rees \cite{rees}, 
a ``dark age''. 
The primordial radiation then cooled below 3000 K, shifting first into the 
infrared and then into the radio. We understand the microphysics of the 
post-recombination universe well. The fractional ionization freezed out 
to the value $10^{-5}\Omega_M/(h\Omega_b)$:\footnote{Throughout 
this talk I will denote with 
$\Omega_M, \Omega_\Lambda,$ and $\Omega_b$ the matter, vacuum, and baryon
density parameters today, and with $h$ the dimensionless Hubble constant,
$h=H_0/100\,\kmsmpc$.}\, these residual 
electrons were enough to keep the matter in thermal equilibrium with the 
radiation via Compton scattering until a thermalization redshift 
$z_t\simeq 800 (\Omega_bh^2)^{2/5}\simeq 160$, i.e. well after the universe
became transparent \cite{peebles}.
Thereafter, the matter temperature dropped as $(1+z)^2$ due to adiabatic 
expansion (Fig. 1) until primordial inhomogeneities in the density field
evolved into the non-linear regime.
\begin{figure}[b]
\begin{center}
\includegraphics[width=.45\textwidth]{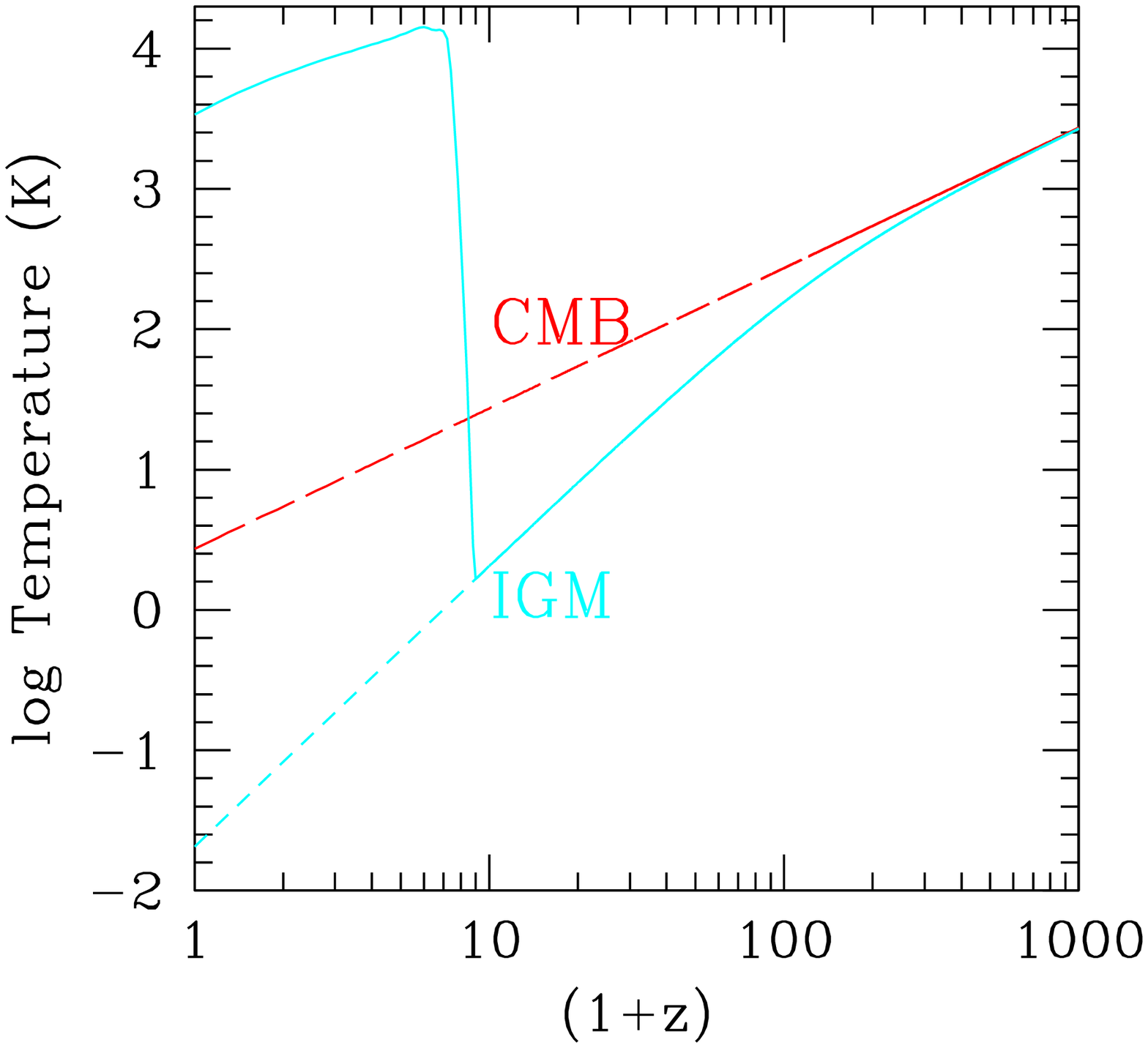}
\includegraphics[width=.45\textwidth]{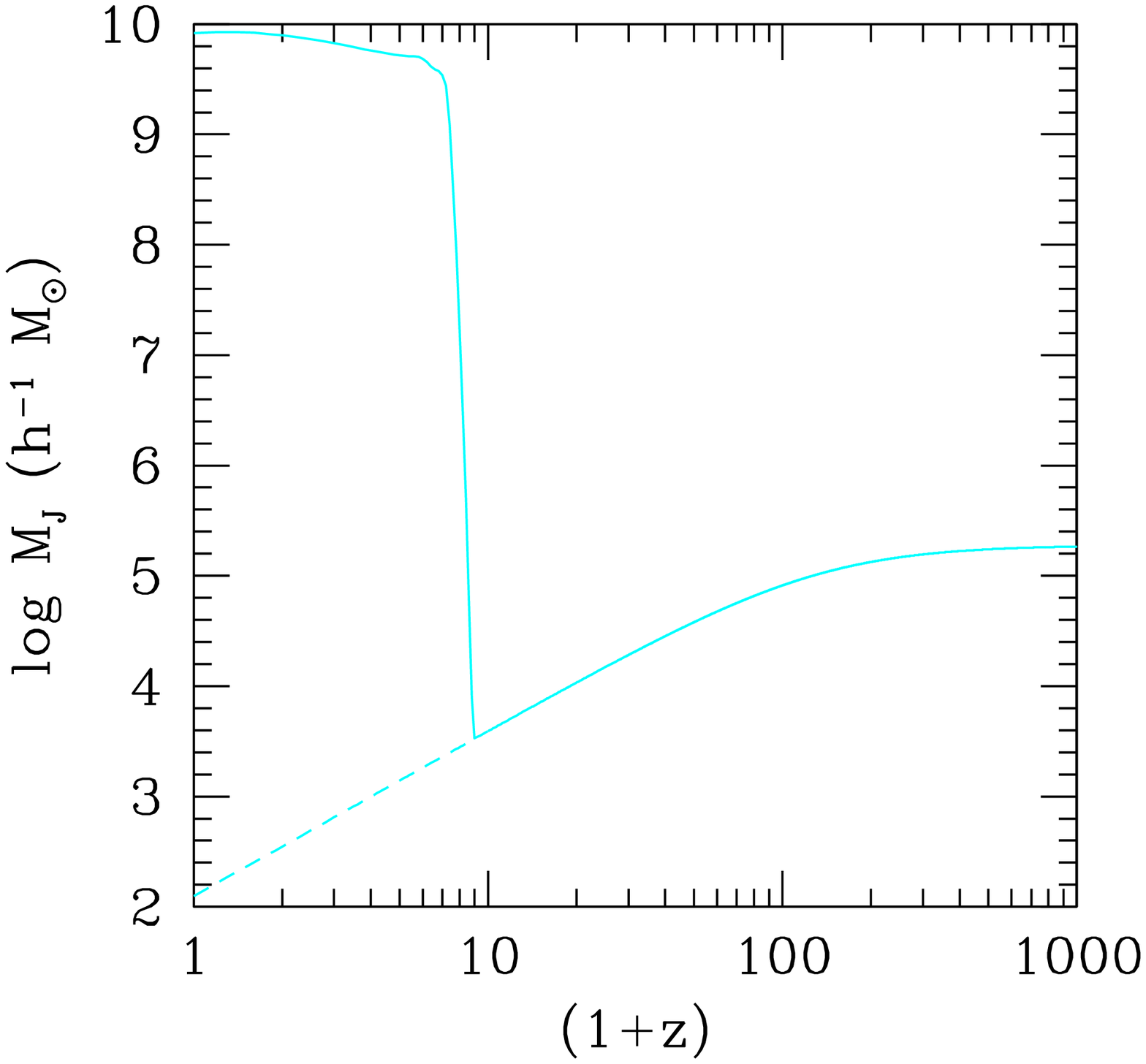}
\end{center}
\vspace{0.3cm}
\caption[]{\footnotesize {\it Left:} Evolution of the radiation
({\it long-dashed line}, labeled CMB) and gas ({\it solid line}, labeled
IGM) temperatures after recombination. The universe is assumed to be reionized 
by ultraviolet radiation at $1+z=10$. The {\it short-dashed line} is the 
extrapolated gas temperature in the absence of any reheating mechanism.    
{\it Right:} Cosmological (gas $+$ dark matter) Jeans mass. 
}
\end{figure}
The minimum mass scale for the gravitational aggregation of
cold dark matter particles is negligibly small. One of the 
most popular CDM candidates is the neutralino: in neutralino CDM, collisional
damping and free streaming smear out all power of primordial density 
inhomogeneities only below $\sim 10^{-7}\,\msun$ \cite{hofman}.  
Baryons, however, respond to pressure gradients and do not fall into dark 
matter clumps below the cosmological Jeans mass (in linear
theory this is the minimum mass-scale of a perturbation where gravity
overcomes pressure) 
\begin{equation}
M_J={\pi \bar\rho\over 6}\left({5\pi kT\over 3G\bar\rho m_p\mu}\right)^{3/2}
\approx 3\times 10^4\,h^{-1}\,\msun (aT/\mu)^{3/2}\Omega_M^{-1/2}. 
\end{equation}
Here $a=(1+z)^{-1}$ is the scale factor, $\bar \rho$ the total mass 
density, $\mu$ the mean molecular weight, and $T$ the gas temperature. 
The evolution of $M_J$ is shown in Fig. 1. In    
the post-recombination universe, the baryon-electron gas is thermally coupled
to the CMB, $T\propto a^{-1}$, and the Jeans mass is independent of redshift
and comparable to the mass of globular clusters, $M_J\approx 10^5\,h^{-1}\,
\msun$. For $z<z_t$, the temperature of the baryons drops as $T
\propto a^{-2}$, and the Jeans mass decreases with time, $M_J\propto a^{-3/2}$.
This trend is reversed by the reheating of the IGM. The energy released 
by the first collapsed objects drives the Jeans mass up to galaxy scales
(Fig. 1): previously growing density perturbations decay as their mass drops
below the new Jeans mass. Photoionization by the ultraviolet radiation 
from the first stars and quasars will heat the IGM to temperatures 
of $\approx 10^4\,$K (corresponding to a Jeans mass $M_J\approx 4\times 
10^9\,h^{-1}\,\msun$ at $1+z=10$),
suppressing gas infall into low mass halos and preventing new (dwarf) 
galaxies from forming.\footnote{It has been pointed out by \cite{gnhui} 
that,
when the Jeans mass itself varies with time, linear gas fluctuations tend to
be smoothed on a (filtering) scale that depends on the full thermal 
history of the gas instead of the instantaneous value of the sound speed:
after reheating, this filtering scale is actually smaller than the 
Jeans scale. Numerical simulations of cosmological reionization show that 
the characteristic suppression mass is typically lower than the 
linear-theory Jeans mass \cite{gnedin}.}   

\section{The epoch of reionization}

Hierarchical clustering theories provide a well-defined
cosmological framework in which the history of baryonic material can
in principle be tracked through the development of cosmic structure. 
Probing the reionization epoch may then provide a means for constraining 
competing models for the formation of cosmic structures. For example, 
popular modifications of the CDM paradigm that attempt to improve over 
CDM by 
suppressing the primordial power-spectrum on small scales, like warm dark 
matter (WDM), are known to reduce the number of collapsed halos at high 
redshifts and make it more difficult to reionize the universe 
\cite{barkana}. 

In practice, we are unable to predict when reionization actually 
occurred. 
While N-body$+$hydrodynamical simulations have convincingly shown that 
intergalactic gas is expected to fragment into structures at early times in 
CDM cosmogonies, the same simulations are much less able to predict the 
efficiency with which the first gravitationally collapsed objects lit up the 
universe at the end of the dark age. The crucial processes of star 
formation, 
feedback (e.g. the effect of the heat input from the first generation of 
sources on later ones), and assembly of massive black holes in the nuclei of
galaxies are poorly understood. Consider the following illustrative 
example:
  
Hydrogen photoionization requires more than one photon above 13.6 eV 
per hydrogen atom: of order $t/\bar t_{\rm rec}\sim 10$ (where 
$\bar t_{\rm rec}$ is the volume-averaged hydrogen recombination 
timescale) extra photons 
appear to be needed to keep the gas in overdense regions and filaments 
ionized against radiative recombinations \cite{gnedin2}. 
A `typical' stellar population produces during its lifetime
about 4000 Lyman continuum (ionizing) photons per stellar proton.
A fraction $f\sim 0.25$\% of cosmic baryons must then
condense into stars to supply the requisite ultraviolet flux. This 
estimate assumes a standard (Salpeter) initial mass function (IMF), 
which determines the relative abundances of hot, high mass stars 
versus cold, low mass ones. 

The very first generation of stars 
(`Population III') must have formed, however, out of unmagnetized 
metal-free gas: numerical simulations of the fragmentation of pure
H and He molecular clouds \cite{bromm} \cite{abel} have shown that these 
characteristics likely led to a `top-heavy' IMF biased towards very 
massive 
stars (VMSs, i.e. stars a few hundred times more massive than the Sun), 
quite different from the present-day Galactic case. Metal-free VMSs
emit about $10^5$ Lyman continuum photons per stellar baryon \cite{bromm2},
approximately $25$ times more than a standard stellar 
population. A corresponding smaller fraction of cosmic baryons would have 
to collapse then into VMSs to reionize the universe, $f\sim 10^{-4}$.
There are of course further complications. Since, at zero 
metallicity, mass loss through radiatively-driven stellar winds is 
expected to be negligible \cite{kudri}, Population  III stars may 
actually die 
losing only a small fraction of their mass. If they retain their large 
mass until death, VMSs with masses $100\lta m\lta 250\,\msun$ will
encounter the electron-positron pair instability and disappear in a 
giant nuclear-powered explosion \cite{fryer}, leaving no compact 
remnants and polluting the universe with the first heavy elements 
\cite{schneider}. In still heavier stars, however, oxygen and
silicon burning is unable to drive an explosion, and complete collapse
to a black hole will occur instead \cite{bond}. 
Thin disk accretion onto a Schwarzschild black hole releases about 
50 MeV per baryon. The conversion of a trace amount of the total 
baryonic mass into early black holes, $f\sim 3\times 10^{-6}$, 
would suffice to reionize the universe.          

Quite apart from the uncertainties in the IMF, it is also unclear 
what fraction of the cold gas will be retained in protogalaxies after
the formation of the first stars (this will affect the global 
efficiency of star formation at these epochs) 
and whether an early input of mechanical energy will play a role in 
determining the thermal and ionization state of the IGM on large scales. 
The same massive stars that emit ultraviolet light also explode as 
supernovae 
(SNe), returning most of the metals to the interstellar medium of
pregalactic systems and injecting about $10^{51}\,$ergs per event 
in kinetic 
energy. For a standard IMF, one has about one SN every $150\,\msun$ 
of baryons that forms stars. A complex network of feedback 
mechanisms is likely at work in these systems, as the gas in shallow 
potential is more easily blown away \cite{dekel},
thereby quenching star formation. Furthermore, as the 
blastwaves produced by supernova explosions sweep the surrounding 
intergalactic gas and enrich it with newly formed heavy elements
\cite{mfr}, they may inhibit the formation of surrounding low-mass 
galaxies due to `baryonic stripping' \cite{scanna}, and drive vast 
portions of the IGM to a significantly higher adiabat than expected 
from photoionization.

\section{Detecting protogalaxies before reionization breakthrough}

By a redshift of 7 no large galaxies should have assembled yet,
but hierarchical clustering theories predict the existence of a 
large population of subgalactic fragments with masses comparable 
to present-day dwarf ellipticals. What are the prospects for 
detecting these early star-forming systems before reionization  
breakthrough? Prior to complete reionization at redshift $z_r$, sources 
of ultraviolet radiation will be seen behind a large column of 
intervening gas that is still neutral. Photons with energies between 
10.2 and 13.6 eV will propagate freely into the IGM until, 
redshifted by the Hubble expansion, they will ultimately reach 
the \lya transition energy of 10.2 eV, resonantly scatter off neutral 
hydrogen, and be removed from the line-of-sight. The spectrum of a 
source at $z>z_r$ should then show a Gunn-Peterson \cite{gp}
absorption trough at wavelengths shorter than $\lambda_\alpha\,
(1+z)$, where $\lambda_\alpha=1216\,$\AA. In fact, the 
Gunn-Peterson optical depth would be so large that the damping wing
of this absorption trough would spill over into the red of \Lya 
\cite{miralda}. This characteristic feature extends for more than 
$1500\,\kms$ to the red of the resonance, and may significantly suppress 
the \lya emission line in the spectra of the first generation of 
cosmic sources. A disappearing \Lya could then be used 
in principle as a flag of the observation of the universe before 
reionization.

\begin{figure}[b]
\begin{center}
\includegraphics[width=.45\textwidth]{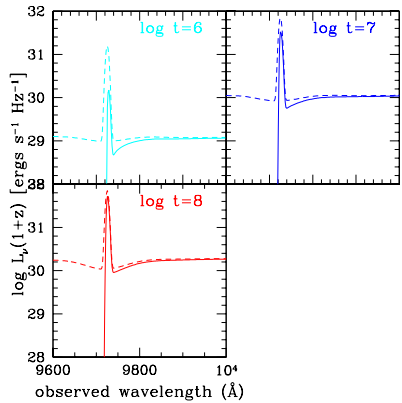}
\end{center}
\vspace{0.3cm}
\caption[]{\footnotesize 
{\it Dashed lines}: Synthetic ultraviolet intrinsic spectra of a 
luminous galaxy at $z=7$ as a function of age $t$ (in yr). The 
stellar population is formed at a constant rate of $\dot M=10\,\sfr$,
with a Salpeter IMF and 
metallicity $Z=0.05\,$Z$_\odot$. The \Lya line is computed in 
Case B recombination assuming an escape fraction of Lyman continuum
photons of 50\% and a Doppler width of $150\,\kms$. 
{\it Solid lines}: Same spectra transmitted through a uniform IGM, 
assuming the source is being observed prior to the reionization epoch. 
}
\end{figure}

In practice, to generate a \Lya emission line by recombination
these objects have to produce ionizing photons. If the escape fraction
for Lyman continuum photons into the intergalactic space is significant,
a photoionized region of IGM will surround each individual source 
\cite{madres}, increasing the transmission of photons redward of 
the \lya resonance. 
If the source lifetime is shorter than the cosmic expansion and gas 
recombination timescales, the volume of ionized gas
will be proportional to the total number of photons produced above 
13.6 eV that leak into intergalactic space; the effect of this 
local photoionization is to greatly
reduce the scattering opacity between the redshift of the source 
and the boundary of its \HII region. In Fig. 2 I have plotted the 
expected ultraviolet spectrum and \lya line profile for a luminous star-forming galaxy 
observed at $z=7$ before reionization breakthrough (assumed to occur at 
$z_r=6$), as a function of stellar age. 
The suppression effect of the \Lya emission line by the red wing of 
the Gunn-Peterson trough, while clearly visible at early times (i.e.
in the absence of a local \HII region, top-left panel), weakens 
significantly with time as the size of the photoionized zone increases.
In this figure the escape fraction of ionizing photons into the IGM 
was assumed to be 50\%. In general (depending on the intrinsic line 
width), the emission line appears to remain observable at late 
times if the escape fraction of Lyman continuum photons exceeds 5--10\% 
(see also \cite{haiman}). 

This should be taken just as an illustrative example, as
in some numerical simulations \cite{gnedin2} reionization was
already well in progress prior to redshift 6, and the form of the damping
profile would be different in the case of patchy ionization along the 
line of sight. Yet, the figure shows that the detection of \lya emission
in the spectra of luminous galaxies cannot then be used, 
by itself, as a constraint on the reionization epoch. 
Even before reionization, the line is unlikely to be obscured  
from view; early star-forming protogalaxies may still be best detected
in the near-IR via their \Lya emission.

\section{First light at 21 cm}

Recent progress with cosmological hydro-simulations based on hierarchical 
structure formation models has led to important insight into the topology
of intergalactic baryons. According to these
calculations, a truely inter- and proto-galactic medium 
collapses under the influence
of dark matter gravity into flattened and filamentary structures,
which are seen in absorption against background quasi stellar sources.
Ballon experiments have produced convincing detections of fluctuations 
in the
temperature of the microwave background radiation on angular scales of
a few arcminutes, providing a direct link between the state of the
universe at recombination and the high contrast structures visible at
later times.

Prior to the epoch of full reionization, the intergalactic medium and 
gravitationally collapsed systems may be detectable in emission or 
absorption against the CMB at the frequency corresponding to the 
redshifted 21 cm line 
(associated with the spin-flip transition from the triplet to the singlet 
state of neutral hydrogen.) In emission, the contribution of a patch of 
neutral IGM with overdensity  $\delta$ to the background spectrum 
at $21\,(1+z)$ cm amounts to a brightness temperature at Earth of
\begin{equation}
\delta T_b\simeq 0.02\,{\rm K}~h^{-1}(1+\delta)\left({\Omega_bh^2\over 0.02}\right)
\left[\left({1+z\over 10}\right)\left({0.3\over \Omega_M}\right)\right]^{1/2},
\end{equation}
very small compared to the CMB. Nonetheless, 21 cm spectral features
are expected to display structure in redshift space as well as angular structure 
due to inhomogeneities in the gas density field, hydrogen ionized 
fraction, and spin temperature.  Several different signatures have been 
investigated in the recent literature: (a) the fluctuations in the 
redshifted 21 cm emission induced by the gas density inhomogeneities 
that develop at early times in CDM-dominated cosmologies \cite{mmr}
\cite{tozzi} and by virialized ``minihalos''  with virial temperatures 
below the threshold for atomic cooling \cite{iliev}; (b) the global 
feature (`reionization step') in the continuum spectrum of the radio 
sky that may mark the abrupt overlapping phase of individual 
intergalactic \HII regions \cite{shaver};
(c) and the 21 cm narrow lines generated in absorption against very
high-redshift radio sources by the neutral IGM \cite{carilli} and by intervening
minihalos  and protogalactic disks \cite{furlanetto}.

\begin{figure}[b]
\begin{center}
\hspace{1.cm}
\includegraphics[width=.45\textwidth]{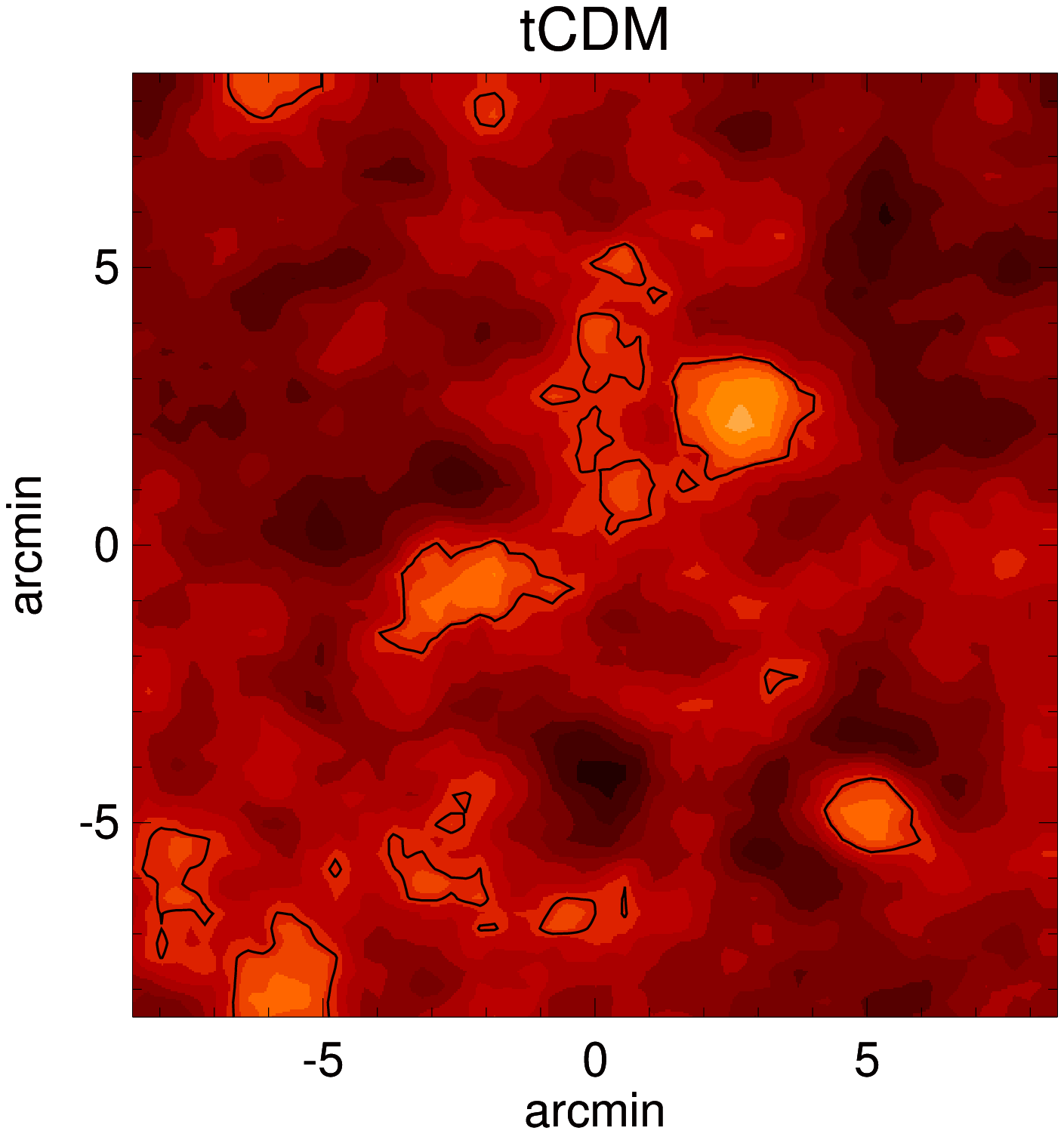}
\includegraphics[width=.45\textwidth]{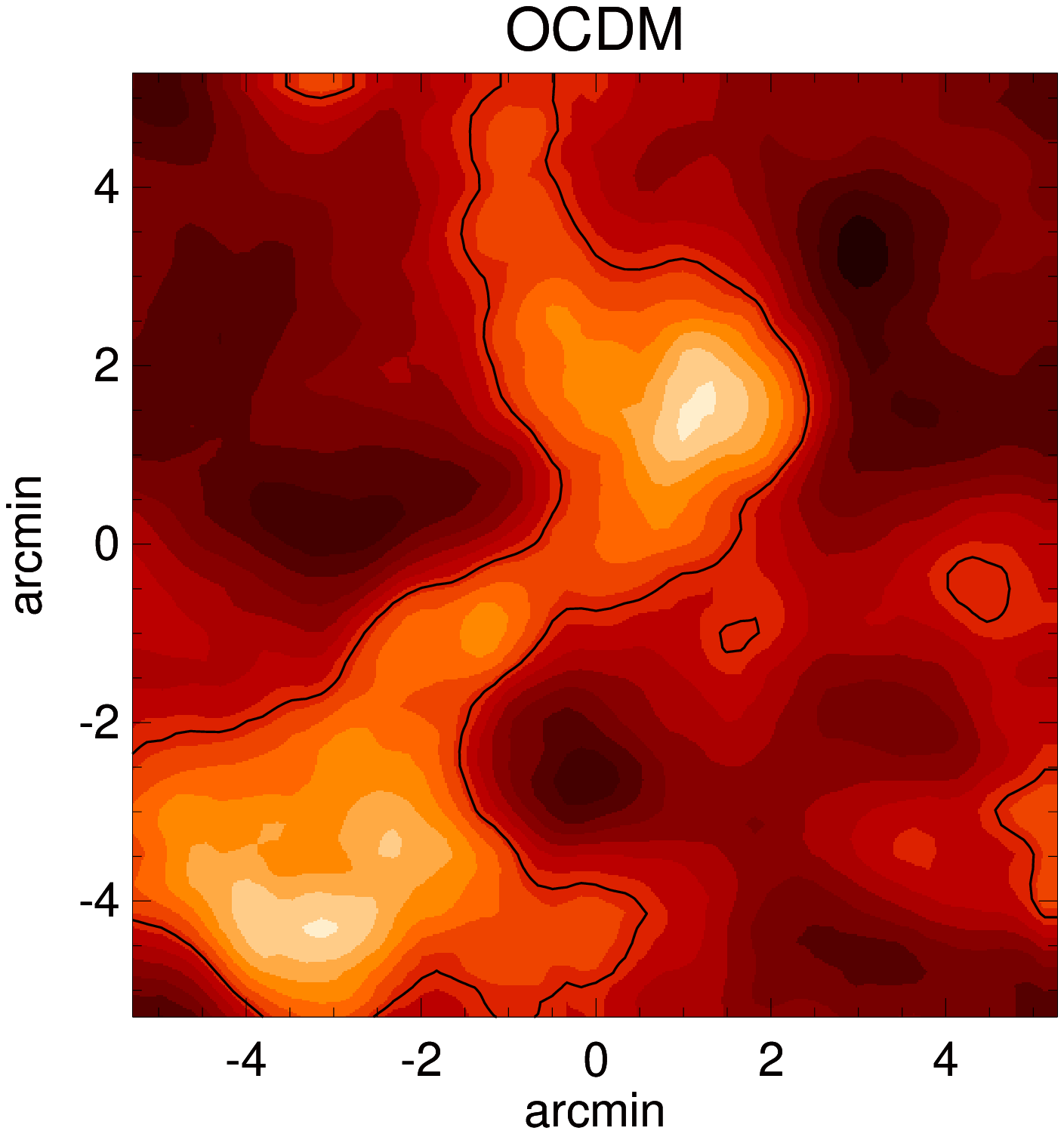}
\end{center}
\vspace{0.3cm}
\caption[]{\footnotesize {\it Left:} Radio map of redshifted 21 cm emission 
against the CMB in a `tilted' CDM (tCDM) cosmology at $z=8.5$. 
Here a collisionless $N$-body simulation with 64$^3$ particles has
been performed with box size $20 h^{-1}$ comoving Mpc, corresponding to 17 (11) 
arcmin in tCDM (OCDM). The baryons are assumed to trace the dark matter
distribution without any biasing, the spin temperature to be much 
greater than the temperature of the CMB everywhere, and the gas to be 
fully neutral. The frequency window is 1 MHz around a central frequency of 
$150$ MHz. The contour levels outline regions with
signal greater than $4\,\mu$Jy per beam. {\it Right:} Same for a open cosmology
(OCDM). Since the growth of density fluctuations ceases early on in an open 
universe (and the power spectrum is normalized to the abundance of 
clusters today), the signal at a given angular size is much larger in OCDM 
than in tCDM at these early epochs \cite{tozzi}.
}
\end{figure}

A quick summary of the physics of 21 cm radiation will illustrate 
the basic ideas and unresolved issues behind these studies.
The emission or absorption of 21 cm photons from a neutral IGM is
governed by the hydrogen spin temperature $T_S$ defined by $n_1/n_0=3
\exp(-T_*/T_S)$,
where $n_0$ and $n_1$ are the number densities of atoms in the
singlet and triplet $n=1$ hyperfine levels
and $T_*=0.068\,$ K is the excitation temperature of the 21 cm transition. 
In the presence of only the CMB radiation with
$T_{\rm CMB}=2.73\,(1+z)\,$K, the spin states will reach thermal equilibrium
with the CMB on a timescale of $T_*/(T_{\rm CMB}A_{10})\approx 3\times10^5\,
(1+z)^{-1}\,$yr ($A_{10}=2.9\times 10^{-15}\,$s$^{-1}$
is the spontaneous decay rate of the hyperfine transition of atomic hydrogen),
and intergalactic \HI will produce neither an absorption nor an emission
signature. A mechanism is required that decouples $T_S$ and $T_{\rm CMB}$,
e.g. by coupling the spin temperature instead to the kinetic temperature
$T_K$ of the gas itself. Two mechanisms are
available, collisions between hydrogen atoms \cite{purcell}
and scattering by \Lya photons \cite{field}. The
collision-induced coupling between the spin and kinetic temperatures
is dominated by the spin-exchange process between the colliding
hydrogen atoms. The rate, however, is too small for realistic IGM
densities at the redshifts of interest, although collisions may be
important in dense regions with $\delta\rho/\rho\gta 30[(1+z)/10]^{-2}$, 
like virialized minihalos.

In the low density IGM, the dominant mechanism is the scattering of 
continuum ultraviolet photons redshifted by the Hubble expansion 
into local \Lya photons. The
many scatterings mix the hyperfine levels of neutral hydrogen in its
ground state via intermediate transitions to the $2p$ state, the
Wouthuysen-Field process. As the neutral IGM is highly opaque to resonant 
scattering, the shape of the continuum radiation spectrum
near \Lya will follow a Boltzmann distribution with a temperature given by
the kinetic temperature of the IGM \cite{field2}. In this case the spin
temperature of neutral hydrogen is a weighted mean between the matter and
CMB temperatures. There exists then a critical value of the background flux of \Lya 
photons
which, if greatly exceeded, would drive the spin temperature away from 
$T_{\rm CMB}$. In \cite{tozzi} we 
used N-body cosmological simulations and, 
assuming a fully neutral medium with $T_S\gg T_{\rm CMB}$, showed that 
prior to 
reionization the same network of sheets and filaments (the `cosmic web') 
that 
gives rise to the \lya forest at $z\sim 3$ should lead to fluctuations in the 
21 cm brightness temperature at higher redshifts (Fig. 3). 
At 150 MHz ($z=8.5$), for
observations with a bandwidth of 1 MHz, the root mean square fluctuations
should be $\sim 10\,$ mK at $1'$, decreasing with scale.
Because of the smoothness of the CMB sky, fluctuations in the 21 cm
radiation will dominate the CMB fluctuations by about 2 orders of 
magnitude on arcmin scales. 

While the microphysics is well understood, our understanding of the 
astrophysics of 21 cm tomography is still poor. As mentioned above,
it is the presence of a sufficient flux of \Lya photons which renders the 
neutral IGM `visible'. Without heating sources, the adiabatic expansion of the
universe will lower the kinetic temperature of the gas well below that of the
CMB, and the IGM will be detectable through its absorption. If there are
sources of radiation that preheat the IGM, it may be possible to detect it 
in emission instead.
The energetic demand for heating the IGM above the CMB temperature is meager,
only $\sim 0.004\,$ eV per particle at $z\sim 10$. Consequently, even
relatively inefficient heating mechanisms may be important warming sources
well before the universe was actually reionized.
Perhaps more importantly, prior to full reionization the 
IGM will be a mixture of neutral, partially ionized,
and fully ionized structures: low-density regions will be fully ionized first, 
followed by regions with higher and higher densities. Radio maps at 
$21\,(1+z)\,$ cm  
will show a patchwork of emission/absorption signals from \HI zones
modulated by \HII regions where no signal is detectable against the CMB.

\section{Summary}

Despite much recent progress in our understanding of the formation 
of early cosmic structure and the high-redshift universe, the end of 
the cosmic ``dark ages'' remains one of missing link in galaxy
formation and evolution studies. We are left very uncertain about the 
whole era from $10^7$ to $10^9$ yr -- the epoch of the 
first galaxies, stars, supernovae, and massive black holes. 
Some of the issues
discussed above are likely to remain a topic of lively controversy
until the launch of the {\it Next Generation Space Telescope}, ideally suited 
to image the earliest generation of stars in the universe. 
If the first massive black holes form in pregalactic systems at very high 
redshifts, they will be incorporated through a series of mergers into larger and 
larger halos, sink to the center owing to dynamical friction, 
accrete a fraction of the gas in the merger remnant to become 
supermassive, and form a binary system \cite{volonteri}. Their coalescence 
would be signalled by the emission of low-frequency gravitational waves
detectable by the planned {\it Laser Interferometer Space Antenna}. The search 
at 21 cm for the epoch of first light has 
become one of the main science drivers of the {\it LOw Frequency ARray}.
While remaining 
an extremely challenging project due to foreground contamination from 
extragalactic radio sources  \cite{dimatteo}, the detection and 
imaging of these small-scale structures with {\it LOFAR} is a tantalizing 
possibility within range of the thermal noise of the array. 

\bigskip\bigskip

I would like to thank all my collaborators for discussions on 
the topics described here, and Martin Rees for many insightful 
suggestions. Support 
for this work was provided by NASA through grant NAG5-11513 and by 
NSF grant AST-0205738.

\end{document}